%% file: main.tex
\documentclass[submission,copyright,creativecommons]{eptcs}
\providecommand{\event}{ML 2018}  
\usepackage{breakurl}             
\usepackage{underscore}           

\usepackage{marginnote}
\usepackage{booktabs} 
\usepackage{amssymb}
\usepackage{amsmath}
\usepackage{mathrsfs}
\usepackage{mathtools}
\usepackage{multirow}
\usepackage{listings}
\usepackage{indentfirst}
\usepackage{verbatim}
\usepackage{amsmath, amssymb}
\usepackage{graphicx}
\usepackage{xcolor}
\usepackage{url}
\usepackage{stmaryrd}
\usepackage{xspace}
\usepackage{comment}
\usepackage{wrapfig}
\usepackage[caption=false]{subfig}
\usepackage{placeins}
\usepackage{tabularx}
\usepackage{ragged2e}
\usepackage{soul}
\usepackage{csquotes}
\usepackage{inconsolata}

\lstdefinelanguage{ocaml}{
keywords={@type, function, fun, let, in, match, with, when, class, type,
nonrec, object, method, of, rec, repeat, until, while, not, do, done, as, val, inherit, and,
new, module, sig, deriving, datatype, struct, if, then, else, open, private, virtual, include, success, failure,
lazy, assert, true, false, end},
sensitive=true,
commentstyle=\small\itshape\ttfamily,
keywordstyle=\ttfamily\bfseries, 
identifierstyle=\ttfamily,
basewidth={0.5em,0.5em},
columns=fixed,
fontadjust=true,
literate={->}{{$\to$}}3 {===}{{$\equiv$}}1 {=/=}{{$\not\equiv$}}1 {|>}{{$\triangleright$}}3 {\\/}{{$\vee$}}2 {/\\}{{$\wedge$}}2 {>=}{{$\ge$}}1 {<=}{{$\le$}} 1,
morecomment=[s]{(*}{*)}
}

\newcommand{\cd}[1]{\texttt{#1}}
\newcommand{\inbr}[1]{\left<#1\right>}
\title{Generic Programming with Combinators and Objects\thanks{This work was partially supported by the grant 18-01-00380 from the Russian Foundation for Basic Research.}}

\author{Dmitry Kosarev
  \institute{St. Petersburg State University\\
    JetBrains Research \\
    St. Petersburg, Russia}
\email{Dmitrii.Kosarev@protonmail.ch}
\and
Dmitry Boulytchev
\institute{St. Petersburg State University\\
  JetBrains Research \\
  St. Petersburg, Russia}
\email{dboulytchev@math.spbu.ru}
}

\def\titlerunning{Generic Programming with Combinators and Objects}
\def\authorrunning{D.Kosarev, D.Boulytchev}
\begin{document}
\maketitle

\begin{abstract}
  We present a generic programming framework for \textsc{OCaml} which makes it possible to implement extensible
  transformations for a large scale of type definitions. Our framework makes use of object-oriented features
  of \textsc{OCaml}, utilising late binding to override the default behaviour of generated transformations. The
  support for polymorphic variant types complements the ability to describe composable data types with the
  ability to implement composable transformations.
\end{abstract}

\input{intro.tex}

\input{exposition.tex}
\input{impl.tex}
\input{examples.tex}
\input{related.tex}
\input{conclusion.tex}

\nocite{*}
\bibliographystyle{eptcs}
\bibliography{main}
\end{document}

%% file: intro.tex
\section{Introduction}

Frederic Brooks in his seminal book on software engineering ``The Mythical Man-Month''~\cite{MMM} has characterised the essence of programming with the following words:

\blockquote{``The programmer, like the poet, works only slightly removed from pure thought-stuff. He builds his castles in the air, from air, creating by exertion of the imagination. Few media of
creation are so flexible, so easy to polish and rework, so readily capable of realising grand conceptual structures. (As we shall see later, this very tractability has its own problems.)''}

Indeed, the virtuality of programs and flexibility of their representation call for structuring; the lack of proper structure easily leads to disastrous consequences
(as it happened to some real-world industrial projects in the past). One of commonly used ways to bring a
structure in software are \emph{data types}. Data types allow to describe the properties of data, what can and cannot be done, and to some extent they prescribe
the semantics to data structures. Being kept in runtime, data types make it possible to implement meta-transformations by analysing types (\emph{introspection})
or even creating new types on the fly (\emph{reflection}).

However, in statically typed languages, as a rule, types are completely erased after the compilation and do not retained in runtime. This has a huge advantage over
dynamic typing as, first, programs do not need to inspect types at runtime anymore and, second, a whole class of bad runtime behaviours~--- type errors~---
is eliminated. The other side of the coin, however, is that now some transformations, which in untyped languages can be implemented ``once and for all'',
can not be typed and have to be re-implemented for each type of interest. One way to overcome this deficiency is to develop a more powerful type system in
which more functions can be typed; as an example we may mention the support for \emph{ad-hoc} polymorphism in \textsc{Haskell} in the forms of type
classes~\cite{TypeClasses} and type families~\cite{TypeFamilies}. However, due to the totality of type checking and fundamental undecidability results there
will always be some ``good'' programs which cannot be typed. Another approach, \emph{datatype-generic programming}~\cite{DGP}, is aimed at developing techniques for
implementation of practically important families of type-indexed functions using existing language features. For example, types can be encoded in a substrate
language~\cite{Hinze,InstantGenerics,GenericOCaml}, or a part of type information can be saved for runtime~\cite{SYB,SYBOCaml}, or generic functions for a given
type can be generated at compile-time automatically~\cite{Yallop,PPXLib}. The two approaches we mentioned are in fact complementary~--- the more powerful
type system is the more means for datatype-generic programming the language can incorporate natively. For example, parametric polymorphism makes it possible
to natively express many generic functions like length of list of arbitrary elements, etc.

We present a generic programming library \textsc{GT}\footnote{\url{https://github.com/kakadu/GT/tree/ppx}} (\emph{Generic Transformers}), which has been in an
active development and use since 2014. One of the important observations, which motivated the development of our framework, was that many generic functions
can be considered as a modifications of some other generic functions. While our approach is generative~--- we generate generic functionality from type definitions~---
it also makes possible for end users to easily derive variants of generated functions by redefining some parts of their functionality. This is achieved using
a method-per-constructor encoding for concrete transformations, which resembles the approach of object algebras~\cite{ObjectAlgebras}.

The main properties of our solution are as follows:

\begin{itemize}
\item each transformation is expressed in terms of a \emph{traversal function} and a \emph{transformation object}, which encapsulate the ``interesting part''
  of the transformation;
\item the traversal function is unique for given type and all transformation objects for the type are instances of a unique class;
\item both the traversal function and the class are generated from type definition; we support regular ADTs, structures, polymorphic variants and predefined types;
\item we provide a number of plugins which generate practically important transformations in the form of concrete transformation classes;
\item the plugin system is extensible: end users can implement their own plugins.
\end{itemize}

The library we present is an inheritor of our earlier work~\cite{SYBOCaml} on implementation of ``Scrap Your Boilerplate'' approach~\cite{SYB,SYB1,SYB2}. However,
our experience has shown, that the expressivity and extensibility of SYB is insufficient; in addition the uniform transformations, based solely on type discrimination, turned out to be
inconvenient to use. Our idea initially was to combine combinator and object-oriented approaches~--- the former would provide means for parameterization, while the
latter~--- for extensibility via late binding utilisation. This idea in the form of a certain design pattern was successfully evaluated~\cite{SCICO} and then reified
in a library and a syntax extension~\cite{TransformationObjects}. Our follow-up experience with the library~\cite{OCanren} has (once again) shown some flaws in the
implementation. The version we present here is almost a complete re-implementation with these flaws fixed.

The rest of the paper is organised as follows. In the next section we give an informal, example-driven exposition of
the approach we use. Then we describe the implementation in more details, highlighting some aspects which we find important or
interesting. In the following section we consider some examples, implemented with the aid of the library. Related works
section observes the relevant approaches and frameworks and compares them to ours. The final section discusses the directions for
future development.


%% file: exposition.tex
\section{Exposition}
\label{expo}

In this section we gradually unfold the approach we propose using a number of examples; while this exposition lacks many concrete details and can not be used as a
precise reference, it presents the main ``ingredients'' of the solution and motivation which has drove us to identify them. From now on we use the following
convention: we denote $\inbr{\dots}$ the representation of a certain notion in the concrete syntax of \textsc{OCaml}. For example, ``$\inbr{f_t}$`` is an encoding of instance
of type-indexed function ``$f$'' for a type `'$t$''. In the concrete syntax it may be expressed as ``\lstinline{f_t}'' but for now we would refrain from specifying
the exact form.

We start from a simple example. Let us have the following type definition for arithmetic expressions:

\begin{lstlisting}
   type expr =
   | Const of int
   | Var   of string
   | Binop of string * expr * expr
\end{lstlisting}

Recursive function ``$\inbr{show_{expr}}$'' (the first evident candidate for generic implementation) converts an expression into its string representation: 

\begin{lstlisting}
   let rec $\inbr{show_{expr}}$ = function
   | Const  n        -> "Const " ^ string_of_int n
   | Var    x        -> "Var " ^ x
   | Binop (o, l, r) ->
      Printf.sprintf "Binop (%S, %s, %s)" o ($\inbr{show_{expr}}$ l) ($\inbr{show_{expr}}$ r)
\end{lstlisting}

The representation, which ``$\inbr{show_{expr}}$'' provides, preserves the names of constructors; this can be convenient for debugging or
serialisation purposes. However, as a rule, an alternative~--- \emph{pretty-printed}~--- representation is desirable as well. In this
representation an expression is shown in its ``natural syntax'' with infix operators and no constructor names, where brackets are
inserted only when they are really needed. Of course, implementing pretty-printer is easy:

\begin{lstlisting}
   let $\inbr{pretty_{expr}}$ e =
     let rec pretty_prio p = function
     | Const  n        -> string_of_int n
     | Var    x        -> x
     | Binop (o, l, r) ->
        let po = prio o in
        (if po <= p then br else id) @@
        pretty_prio po l ^ " " ^ o ^ " " ^ pretty_prio po r
     in
     pretty_prio min_int e
\end{lstlisting}

Here we make use of functions ``\lstinline{prio}'', ``\lstinline{br}'' and ``\lstinline{id}'', defined elsewhere. ``\lstinline{prio}''
returns the priority of a binary operator, ``\lstinline{br}'' puts its parameter in brackets and ``\lstinline{id}''
is identity. The auxiliary function ``\lstinline{pretty_prio}'' takes additional integer parameter, which describes the priority of an enclosing
binary operator (if any). If the priority of current operator is less of equal than that, the expression is taken into brackets (for simplicity we assume all
operators non-associative; the same code skeleton with minor modifications can be used for the associative case as well). On the top level we supply the
smallest representable integer as the priority to make sure no brackets will appear around the top level expression.

The bodies of these two functions have very little in common~--- both return strings, but the second takes additional argument, and all the constructor cases
are essentially different. The only identical thing is pattern matching itself. We can extract the pattern matching into a separate function and parameterise this
function with a set of per-constructor transformations:

\begin{lstlisting}
   let $\inbr{gcata_{expr}}$ $\omega$ $\iota$ = function
   | Const n         -> $\omega$#$\inbr{Const}$ $\iota$ n
   | Var   x         -> $\omega$#$\inbr{Var}$   $\iota$ x
   | Binop (o, l, r) -> $\omega$#$\inbr{Binop}$ $\iota$ o l r
\end{lstlisting}

Here we use object as a natural representation for a set of semantically connected functions. ``$\omega$'' is a \emph{transformation object} with methods corresponding to
the constructors of type ``\lstinline{expr}''; ``$\iota$'' represents the extra parameter which may be used by functions like ``$\inbr{pretty_{expr}}$'' (and safely
ignored by functions like ``$\inbr{show_{expr}}$'').

The initial ``$\inbr{show_{expr}}$'' now can be expressed as follows\footnote{For the sake of brevity we omitted some type annotations, needed for this snippet to type check.}:

\begin{lstlisting}
   let rec $\inbr{show_{expr}}$ e = $\inbr{gcata_{expr}}$
     object
       method $\inbr{Const}$ _ n   = "Const " ^ string_of_int n
       method $\inbr{Var}$  $\enspace$   _ x   = "Var " ^ x
       method $\inbr{Binop}$ _ o l r =
         Printf.sprintf "Binop (%S, %s, %s)" o ($\inbr{show_{expr}}$ l) ($\inbr{show_{expr}}$ r)
     end
     ()
     e
\end{lstlisting}

and, of course, the same is true for $\inbr{pretty_{expr}}$.

We can notice, that both objects, needed to implement these functions, can be instantiated from a common virtual class:

\begin{lstlisting}
   class virtual [$\iota$, $\sigma$] $\inbr{expr}$ =
   object
     method virtual $\inbr{Const}$ : $\iota$ -> int -> $\sigma$
     method virtual $\inbr{Var}\enspace\;\;$ : $\iota$ -> string -> $\sigma$
     method virtual $\inbr{Binop}$ : $\iota$ -> string -> expr -> expr -> $\sigma$  
   end
\end{lstlisting}

A concrete transformation class inherits from this common ancestor; as we have to make recursive calls to the transformation 
itself we parameterise the class by the self-transforming function ``\lstinline{fself}'' (\emph{open recursion}). The decision to
use open recursion is vital for the support of polymorphic variant types and mutual recursion. Now we can implement, say, pretty-printing
in isolation (not within the pretty-printing function, note the usage of ``\lstinline{fself}''):

\begin{lstlisting}
   class $\inbr{pretty_{expr}}$ (fself : $\iota$ -> expr -> $\sigma$) =
   object inherit [int, string] $\inbr{expr}$ 
     method $\inbr{Const}$ p n = string_of_int n
     method $\inbr{Var}$ p x = x
     method $\inbr{Binop}$ p o l r =
       let po = prio o in
       (if po <= p then fun s -> "(" ^ s ^ ")" else fun s -> s) @@
       fself po l ^ " " ^ o ^ " " ^ fself po r
   end
\end{lstlisting}

The pretty-printing function itself can now be easily expressed using this class and the generic transformation\footnote{As function and class names reside in
  different namespaces in \textsc{OCaml}, we use the same name for both concrete transformation class and transformation function.}:

\begin{lstlisting}
   let $\inbr{pretty_{expr}}$ e =
     let rec pretty_prio p e = $\inbr{gcata_{expr}}$ (new $\inbr{pretty_{expr}}$ pretty_prio) p e in
     pretty_prio min_int e
\end{lstlisting}

Finally, we can avoid using the nested function definition by tying the recursive knot with the fix point combinator ``\lstinline{fix}'':

\begin{lstlisting}
   let $\inbr{pretty_{expr}}$ e =
     fix (fun fself p e -> $\inbr{gcata_{expr}}$ (new $\inbr{pretty_{expr}}$ fself) p e) min_int e
\end{lstlisting}

During this demonstration we managed to extract two common features for two essentially different transformations: a generic traversal (``$\inbr{gcata_{expr}}$'')
and a virtual class (``$\inbr{expr}$'') to represent all transformations as its instances. But, did it worth trying? In fact in this concrete example we achieved a
very little code reuse at the price of introducing a number of extra abstractions; actually, the size of code we came up with is \emph{larger} than the initial one.

We argue that in this particular case the transformations were not general enough. In order to justify our approach we consider another, more optimistic scenario. It is
well-known, that many transformations can be represented (and for a good reason) using \emph{catamorphisms}, or ``folds''~\cite{Fold,Bananas,CalculatingFP}. Technically, to
implement regular catamorphism we would need to abstract the type ``\lstinline{expr}'' of itself to make it a proper functor, but for now we stick with a more
lightweight version:

\begin{lstlisting}
   class [$\iota$] $\inbr{fold_{expr}}$ (fself : $\iota$ -> expr -> $\iota$) =
   object inherit [$\iota$, $\iota$] $\inbr{expr}$ 
     method $\inbr{Const}$ i n = i
     method $\inbr{Var}$ i x = i
     method $\inbr{Binop}$ i o l r = fself (fself i l) r
   end
\end{lstlisting}

This implementation simply threads the argument ``\lstinline{i}'' through all nodes of an expression and returns it unchanged. This seems pretty useless at a first
glance. However, if we modify this default behaviour a little, we can obtain something useful:

\begin{lstlisting}
   let fv e =
     fix (fun fself i e ->
            $\inbr{gcata_{expr}}$ (object inherit [string list] $\inbr{fold_{expr}}$ fself
                         method $\inbr{Var}$ i x = x :: i
                       end) i e
         ) [] e
\end{lstlisting}

This function calculates the list of all free variables in an expression (as there can be no binders this is simply the list of all variables). Immediate object we
construct here inherits from the ``useless'' ``$\inbr{fold_{expr}}$'' and redefines only one method~--- for variables. All other code makes exactly what we need~---
``$\inbr{gcata_{expr}}$'' traverses the expression, and all other methods of transformation object accurately pass the list of variables through. So, we indeed
managed to implement some interesting transformation with a very small modification of existing code (provided that ``$\inbr{fold_{expr}}$'' class was already supplied).
To avoid the impression that we carefully prepared everything to implement this particular example we can show another one:

\begin{lstlisting}
   let height e =
     fix (fun fself i e ->
            $\inbr{gcata_{expr}}$ (object inherit [int] $\inbr{fold_{expr}}$ fself
                         method $\inbr{Binop}$ i _ l r = 1 + max (fself i l) (fself i r) 
                       end) i e
         ) 0 e
\end{lstlisting}

Now we calculated the height of an expression. We used the same ``$\inbr{fold_{expr}}$'' class as a base for another immediate object; we redefined the method for
binary operators, which now calculates the heights of both sub expressions, takes the maximum and adds one. 

Another commonly recognised generic feature is ``map'':

\begin{lstlisting}
   class $\inbr{map_{expr}}$ fself =
   object inherit [unit, expr] $\inbr{expr}$
     method $\inbr{Var}$ _ x = Var x
     method $\inbr{Const}$ _ n = Const n
     method $\inbr{Binop}$ _ o l r = Binop (o, fself () l, fself () r)
   end
\end{lstlisting}

Again, as type ``\lstinline{expr}'' is not a functor, all we can do with ``$\inbr{map_{expr}}$'' is copying. However, by inheriting from it we
can provide more transformations:

\begin{lstlisting}
   class simplify fself =
   object inherit $\inbr{map_{expr}}$ fself
     method $\inbr{Binop}$ _ o l r =
       match fself () l, fself () r with
       | Const l, Const r -> Const ((op o) l r)
       | l      , r       -> Binop (o, l, r)     
   end
\end{lstlisting}

This class performs a constant folding: if both arguments of a binary operator are reduced (by the same transformation) to constants, then in
performs the operation. The function ``\lstinline{op}'' is defined elsewhere; it returns an integer function for evaluating given binary operator. One more:

\begin{lstlisting}
   class substitute fself state =
   object inherit $\inbr{map_{expr}}$ fself
     method $\inbr{Var}$ _ x = Const (state x)  
   end
\end{lstlisting}

This one substitutes variables in an expression with their values in some state, represented as function ``\lstinline{state}''. Two last
classes can be seamlessly combined to construct an evaluator:

\begin{lstlisting}
   class eval fself state =
   object
     inherit substitute fself state
     inherit simplify   fself
   end

   let eval state e =
     fix (fun fself i e -> $\inbr{gcata_{expr}}$ (new eval fself state) i e) () e  
\end{lstlisting}

In all these examples we, starting from some very common generic feature, implemented all needed transformations with a very little efforts (modulo
the verbose \textsc{OCaml} syntax for objects and classes). In each case we needed to override only one method, and we used a single per-type generic
function. On the other hand we dealt with a very simple type~--- for example, it was not even polymorphic, and supporting polymorphism might have
its own issues. In the rest of the paper we show that, indeed, the sketch we presented here can be extended to a generic programming
framework, in which all the components can be synthesised from type definitions. In particular, our approach provides the full support for:

\begin{itemize}
\item Polymorphism.
\item Type constructor application.
\item Mutual recursion. While there is no problem with implementation of hard-coded generic transformations, the implementation of \emph{extensible} ones
  requires extra efforts.
\item Polymorphic variant types. It includes the seamless integration via class inheritance of all features
  for polymorphic variant types when these types are combined into the one.
\item Separate compilation: we can generate code from type definitions for a module separately with no lookup into
  modules this one depends on.
\item Encapsulation: we support module signatures, including abstract and private type declarations. Generic functions, implemented for
  abstract types, can be safely used outside the module, but can be neither modified nor used to ``peep'' at the internal structure of
  the type.  
\end{itemize}

We also address some performance issues~--- as one could notice, in all preceding examples we created a whole bunch of \emph{identical} objects during a
transformation (one per each node of a data structure); as we will see, this can be avoided via memoization. Finally, our framework provides a plugin system which can be
used to generate a number of useful transformations (like ``\lstinline{show}'', ``\lstinline{fold}'' or ``\lstinline{map}''). The plugin system is
extensible as well~--- end users can implement their own plugins with a very little amount of extra effort since a large part of their functionality (the traversal
function and virtual transformation class) is already supplied by the framework. 

%% file: impl.tex
\section{Implementation}

The main components of our solution are syntax extensions (both in terms of \cd{camlp5}~\cite{Camlp5} and \cd{ppxlib}~\cite{PPXLib}), a runtime library and
a plugin system. The syntax extensions process type definitions, attributed by an end user, and generate the following entities:

\begin{itemize}
\item a transformation function (one per a type);
\item a virtual class which is used as a common ancestor for all concrete transformations (one per a type);
\item a number of concrete classes (one per requested plugin);
\item a \emph{typeinfo} structure, which incorporates a type-specific information like the transformation function and a bundle of
  plugin-generated concrete functions, represented as an immediate object.
\end{itemize}

We support the majority of variants in the right-hand side of type definitions with the following limitations:

\begin{itemize}
\item only regular algebraic data types are supported; GADTs are treated as simple algebraic data types;
\item constraints are not taken into account;
\item ``\lstinline{nonrec}'' definitions, object and module types are not supported;
\item extensible datatypes (``\lstinline{...}''/``\lstinline{+=}'') are not supported.
\end{itemize}

For example, for a type ``\lstinline{t}'' with requested plugin ``\lstinline{show}'' the structure with the following skeleton is generated (``$\dots$'' stands for the parts we omit for now):

\begin{figure}[t]
  \center
  \begin{tabular}{l|l}
    \hline
    \multicolumn{2}{c}{\cd{camlp5} version}\\
    \hline
    \lstinline|@type ... $[$ with  $p_1, p_2, \dots$ $]$| & a syntax construct to generate a support for a type \\
                                                         & with plugins $p_1, p_2, \dots$; mutually recursive definitions \\
                                                         & are supported \\
    \lstinline|@$typ$| & the name for the virtual class for type $typ$ \\
    \lstinline|@$plugin$[$typ$]| & the name for a plugin class for type $typ$ and \\
                                 & plugin $plugin$\\
    \hline
        \multicolumn{2}{c}{\cd{ppxlib} version}\\
    \hline
    \lstinline|type ... = ...|  & a syntax construct to generate a support for a type \\
    \lstinline|and  ... = ...|  & with plugins $p_1, p_2, \dots$ \\
    \lstinline|[@@deriving gt ~options:{ $p_1, p_2, \dots$}]|
  \end{tabular}
  \caption{Extended syntax constructs}
  \label{syntax}
\end{figure}

\begin{lstlisting}
   let $\inbr{gcata_t}$ $\dots$ = $\dots$
   
   class virtual [$\dots$] $\inbr{t}$ =
   object
     $\dots$
   end

   class [$\dots$] $\inbr{show_t}$ $\dots$ =
   object inherit [$\dots$] $\inbr{t}$ $\dots$
     $\dots$
   end

   let t = {
     gcata   = $\inbr{gcata_t}$;
     $\dots$
     plugins = object
                 method show = $\dots$
               end
   }
\end{lstlisting}

Using the typeinfo structure ``\lstinline{t}'' we can mimic the type-indexed nature of concrete transformations:

\begin{lstlisting}
   let transform typeinfo = typeinfo.gcata
   let show      typeinfo = typeinfo.plugins#show
\end{lstlisting}

The function ``\lstinline{transform(t)}'' is a top-level transformation function, which can be instantiated for any supported type ``\lstinline{t}''. On the
Figure~\ref{syntax} we describe the concrete constructs, implemented by the syntax extensions. Note, the concrete way of encoding names
for classes and transformation function (represented above as $\inbr{...}$) is not important as long as \cd{camlp5} is
used since it provides corresponding syntax extensions.

\subsection{Types of Transformations}

The design of the library is based on the idea to describe transformations (e.g. catamorphisms~\cite{Bananas}) in terms of attribute grammars~\cite{AGKnuth,AGSwierstra,ObjectAlgebrasAttribute}.
In short, we consider only the transformations of the following type

\[
\iota \to t \to \sigma
\]

where $t$ is the type of a value to transform, $\iota$ and $\sigma$~--- types for \emph{inherited} and \emph{synthesised} attributes. We do not use attribute
grammars as a mean to describe the algorithmic part of transformations; we only utilise their terminology to describe the types. 

When the type under consideration is parameterised, the transformation becomes parameterised as well. From now on we will use a convention to
denote $\left\{...\right\}$ multiple occurrences of a an entity inside the brackets. Under this convention we may stipulate the generic form of
any transformation, representable with the aid of our library, as

\[
  \left\{\iota_i \to \alpha_i \to \sigma_i\right\}\to\iota \to\left\{\alpha_i\right\}\;t \to \sigma
\]

Here $\iota_i\to\alpha_i\to\sigma_i$ is an argument-transforming function for the type parameter $\alpha_i$. In general the argument-transforming functions operate on
inherited values of different types and return synthesised values of different types. The common ancestor class in turn is massively polymorphic: for an $n$-parametric
type it receives $3(n+1)$ type parameters:

\begin{itemize}
\item a triplet $\iota_i$, $\alpha_i$, $\sigma_i$ for each type parameter $\alpha_i$, where $\iota_i$ and $\sigma_i$ are type variables for inherited and
  synthesised attributes for the transformation of $\alpha_i$;
\item a pair of type variables $\iota$ and $\sigma$ for inherited and synthesised attributes for the type itself;
\item an extra type variable $\epsilon$, which is inferred to ``\lstinline|$\{\alpha_i\}$ t|'' for non-polymorphic variant types and to
  an \emph{open} version ``\lstinline|[> $\{\alpha_i\}$ t]|'' for polymorphic variants (see Section~\ref{pv}).
\end{itemize}

For example, if we have a two-parametric type \lstinline{($\alpha$, $\beta$) t} the head of common ancestor class definition will look like

\begin{lstlisting}
  class virtual [$\iota_\alpha$, $\alpha$, $\sigma_\alpha$, $\iota_\beta$, $\beta$, $\sigma_\beta$, $\iota$, $\epsilon$, $\sigma$] $\inbr{t}$
\end{lstlisting}

The concrete transformations inherit from the common ancestor class and, possibly, instantiate some of its type parameters to a more
specific types. Additionally, concrete classes receive a number of functional arguments:

\begin{itemize}
\item $n$ argument-transforming functions: \lstinline|f$_{\alpha_i}$ : $\iota_i$ -> $\alpha_i$ -> $\sigma_i$|;
\item a function to implement open recursion: \lstinline|fself : $\iota$ -> $\epsilon$ ->  $\sigma$|.
\end{itemize}

For example, for the same type as above and a transformation ``\lstinline{show}'' the header of concrete class looks like

\begin{lstlisting}
  class [$\alpha$, $\beta$, $\epsilon$] $\inbr{show_t}$ 
    (f$_\alpha$ : unit -> $\alpha$ -> string)
    (f$_\beta$ : unit -> $\beta$ -> string)
    (fself : unit -> $\epsilon$ -> string) =
  object inherit [unit, $\alpha$, string, unit, $\beta$, string, unit, $\epsilon$, string] $\inbr{t}$
    $\dots$
  end 
\end{lstlisting}

Note, we maintain these conventions for all types although for some of them some of components are superfluous: for example, ``\lstinline{fself}''
is needed only for recursive types. The explanation for this decision is simple: when we \emph{use} a type we generally do not know its
definition. Thus, in order to support separate compilation the interfaces of all entities we generate must have identical structure.

This scheme of typing and pasteurisation looks quite verbose and cumbersome: there are a lot of type parameters which are quite easy to get a mess with. However, end
users would need to deal with this stuff directly only when they desire to implement a transformation \emph{manually} from scratch by immediately inheriting from the common ancestor class.
In the majority of use cases the transformation is implemented either by customising a certain plugin or using the plugin system. In the first case many
type parameters are already instantiated (for example, for ``\lstinline{show}'' the majority of type parameters are instantiated to ground types), in the
second the plugin system takes care of instantiating the parameters correctly (see Section~\ref{plugins}).

We also need to describe the types for the methods of common ancestor classes. The method for a
constructor ``\lstinline|C of a$_1$ * a$_2$ * ... * a$_k$|'' has the following definition:

\begin{lstlisting}
   method virtual $\inbr{C}$ : $\iota$ -> $\epsilon$ -> a$_1$ -> a$_2$ -> ... -> a$_k$ -> $\sigma$
\end{lstlisting}

Note, the method takes not only inherited attribute and the arguments of corresponding constructor, but the value under transformation itself.

Finally, we describe the type of transformation function. This type is slightly different for polymorphic variant types.

For a non-polymorphic variant type ``\lstinline|$\{\alpha_i\}$ t|'' the transformation function has the following type:

\begin{lstlisting}
   val $\inbr{gcata_t}$ : [$\{\iota_{\alpha_i}$, $\alpha_i$, $\sigma_{\alpha_i}\}$, $\iota$, $\{\alpha_i\}$ t, $\sigma$]#$\inbr{t}$ -> $\iota$ -> $\{\alpha_i\}$ t -> $\sigma$
\end{lstlisting}

Thus, it takes a transformation object (which has a type of properly parameterised \emph{subclass} of the common ancestor class), an inherited
attribute, a value to transform, and returns synthesised attribute. The extra type parameter ``$\epsilon$'' is instantiated to the
type itself. For a polymorphic variant type the extra type parameter is instantiated to the \emph{open} version of the
type (``\lstinline{[> $\{\alpha_i\}$ t]}''). This enables the possibility to apply a transformation function for a type to a transformation object for
another type with more constructors.

\subsection{Fixed Point Combinator and Memoization}
\label{memofix}

In our approach we use open recursion: a class for a concrete transformation takes a function for the same transformation as a parameter. In order
to instantiate this function we have to use a fix point combinator. In this section we consider only a simple fix point combinator for an isolated
type definition; in mutually-recursive case a more elaborated combinator is needed (see Section~\ref{murec}).

We repeat here an example from Section~\ref{expo}:

\begin{lstlisting}
   let $\inbr{pretty_{ex pr}}$ e =
     fix (fun fself p e -> $\inbr{gcata_{expr}}$ (new $\inbr{pretty_{expr}}$ fself) p e) min_int e
\end{lstlisting}

As the lambda argument of ``\lstinline{fix}'' is called each time when ``\lstinline{fself}'' is called (virtually, for each node of
an expression), a new transformation object is created for each node. As all these objects are identical, this can be optimised. 

We memoize the creation of transformation objects using lazy evaluation. For this we abstract the object creation sub expression into a
function which takes ``\lstinline{fself}'' as an argument. Then the implementation of the fix point combinator is as follows:

\begin{lstlisting}
   let fix gcata make_obj $\iota$ x =
     let rec obj = lazy (make_obj fself)
     and fself $\iota$ x = gcata (Lazy.force obj) $\iota$ x in
     fself $\iota$ x
\end{lstlisting}

This combinator can be used for all types and is not generated. Now we can fix a little bit the definition of ``\lstinline{transform}'':

\begin{lstlisting}
   let transform typeinfo = fix typeinfo.gcata
\end{lstlisting}

With this definition an end used does not need to deal with the fix point combinator explicitly anymore:

\begin{lstlisting}
   let $\inbr{show_{expr}}$ e =
     transform(expr) (fun fself -> new $\inbr{show_{expr}}$ fself) () e
\end{lstlisting}

\subsection{The Plugin System}
\label{plugins}

\begin{figure}[t]
  \center
  \small
  \begin{tabular}{ccp{4cm}}
    Name & Type of the transformation & Comment \\[3mm]
    \hline\\
    \lstinline|show| & \lstinline|$\{$ unit -> $\alpha_i$ -> string $\}$ -> unit -> $\{\alpha_i\}$ t -> string| & conversion to a string \\[2mm]
    \lstinline|fmt| & \lstinline|$\{$ formatter -> $\alpha_i$ -> unit $\}$ -> formatter -> $\{\alpha_i\}$ t -> unit| & formatted output using the ``\lstinline|Format|'' module \\[2mm]
    \lstinline|html| & \lstinline|$\{$ unit -> $\alpha_i$ -> HTML.t $\}$ -> unit  -> $\{\alpha_i\}$ t -> HTML.t| & conversion to HTML representation\\[2mm]
    \lstinline|compare| & \lstinline| $\{$ $\alpha_i$ -> $\alpha_i$ -> comparison $\}$ -> $\{\alpha_i\}$ t -> $\{\alpha_i\}$ t -> comparison| & comparison \\[2mm]
    \lstinline|eq| & \lstinline|$\{$ $\alpha_i$ -> $\alpha_i$ -> bool $\}$ -> $\{\alpha_i\}$ t -> $\{\alpha_i\}$ t -> bool| & equality test \\[2mm]
    \lstinline|foldl| & \lstinline |$\{$ $\alpha$ -> $\alpha_i$ -> $\alpha$ $\}$ -> $\alpha$ -> $\{\alpha_i\}$ t -> $\alpha$| & threading an inherited attribute through all the nodes using a top-down traversal \\[2mm]
    \lstinline|foldr| & \lstinline |$\{$ $\alpha$ -> $\alpha_i$ -> $\alpha$ $\}$ -> $\alpha$ -> $\{\alpha_i\}$ t -> $\alpha$| & threading an inherited attribute through all the nodes using a bottom-up traversal \\[2mm]
    \lstinline|gmap| & \lstinline|$\{$ unit -> $\alpha_i$ -> $\beta_i$ $\}$ -> unit -> $\{\alpha_i\}$ t -> $\{\beta_i\}$ t| & a functor
  \end{tabular}
  \caption{The list of predefined plugins}
  \label{listofplugins}
\end{figure}

The default behaviour of our framework is to generate the transformation function, the common ancestor class and the typeinfo structure only. It does not
generate any concrete built-in transformations. All concrete transformations are generated by \emph{plugins}, and the plugin system allows end users to
implement their own. There is a number of predefined plugins (see Figure~\ref{listofplugins}), but none of them receives a special treatment from the
rest of the framework.

Each plugin is implemented as a dynamically-loaded object, and to create a plugin an end user has to properly instantiate a compilation unit using an interface
provided by the framework. The same approach is used in a number of existing frameworks~\cite{PPXLib,Yallop}; however, we claim, that in our case the
implementation of a plugin is much simpler. The reason is that the concrete and generic parts of transformations are properly separated. Thus,
a plugin only instantiates a class, and only a limited assistance from an end-user side is needed. Generally speaking, the following information has to be
provided:

\begin{itemize}
\item What are the types of inherited and synthesised attributes for a given type parameter?
\item What are the types of inherited and synthesised attributes for the type itself?
\item What is the body of the method which transforms given constructor (the arguments of the method and their types are specified by the framework)?
\item What the toplevel method of the typeinfo structure for the plugin is look like?
\end{itemize}

So, there are only a limited number of places where a plugin actually needs to generate a code, and as a rule the generated code is very simple. The
code generation interface the plugin system provides resembles that of \cd{ppxlib} (more precise, \cd{Ast_builder}), which has to be familiar to anyone,
who has ever implemented syntax extensions. In the Section~\ref{pluginExample} we present a complete example of fresh plugin implementation.

\subsection{Mutual Recursion}
\label{murec}

The full support for mutually recursive type definitions requires extra efforts. While, formally, the generation of all needed entities for
mutually recursive definitions can be done in a similar manner as for the isolated case, it would break the extensibility of
transformations. We demonstrate this phenomenon by the following example. Let us have the definition

\begin{lstlisting}
   type expr = $\dots$ | LocalDef of def * expr
   and  def  = Def of string * expr
\end{lstlisting}

where we omitted a non-relevant part (variables, binary operators, etc.) in expression type declaration. It is rather obvious, that generic
transformation functions for both types can be kept as they are; indeed, they only ``outsource'' the transformations to corresponding
methods and do not depend on recursion in type definitions:

\begin{lstlisting}
   let $\inbr{gcata_{expr}}$ $\omega$ $\iota$ = function
   $\dots$
   | LocalDef (d, e) as x -> $\omega$#$\inbr{LocalDef}$ $\iota$ x d e

   let $\inbr{gcata_{def}}$ $\omega$ $\iota$ = function
   | Def (s, e) as x -> $\omega$#$\inbr{Def}$ $\iota$ x s e
\end{lstlisting}

The same is true for the common ancestor classes. However, when we start implementing concrete transformations, we would need to use a transformation
for ``\lstinline{expr}'' inside the class for ``\lstinline{def}'', and vice versa. This can be done with mutually recursive class definitions (we, again,
omit the non relevant parts):

\begin{lstlisting}
   class $\inbr{show_{expr}}$ fself =
   object inherit [unit, _, string] $\inbr{expr}$ fself
     $\dots$
     method $\inbr{LocalDef}$ $\iota$ x d e =
       $\dots$ (fix $\inbr{gcata_{def}}$ (fun fself -> new $\inbr{show_{def}}$ fself) $\dots$) $\dots$
   end
   and $\inbr{show_{def}}$ fself =
   object inherit [unit, _, string] $\inbr{def}$ fself
     method $\inbr{Def}$ $\iota$ x s e =
       $\dots$ (fix $\inbr{gcata_{expr}}$ (fun fself -> new $\inbr{show_{expr}}$ fself) $\dots$) $\dots$
   end
\end{lstlisting}

Note, in both ``\lstinline{fix}'' sub expressions we instantiated \emph{concrete} classes (``$\inbr{show_{def}}$'' and ``$\inbr{show_{expr}}$''). This should
work as expected at the first glance. Strictly speaking, this \emph{concrete} transformation works. But, what happens when we decide to redefine the behaviour of
this default ``$\inbr{show_{expr}}$''? According to our general approach, we would need to inherit from ``$\inbr{show_{expr}}$'', override certain methods and
construct a function using fix point:

\begin{lstlisting}
   class custom_show fself =
   object inherit $\inbr{show_{expr}}$ fself
     method $\inbr{Const}$ $\iota$ x n = "a constant"
   end

   let custom_show e = fix $\inbr{gcata_{expr}}$ (fun fself -> new custom_show fself) () e
\end{lstlisting}

Alas, this won't work as we desire: we did not override the method ``$\inbr{LocalDef}$'', and it still uses the default version for the type ``\lstinline{def}'', which
still uses the default version for the type ``\lstinline{expr}''. Thus, we only redefined the behaviour of default ``$\inbr{show_{expr}}$'' for one component of
mutually recursive type definition~--- the type ``\lstinline{expr}'' as such. All occurrences of ``\lstinline{expr}'' inside other types will still be handled by
the default transformation. In order to make things work as we want we would need to repeat the \emph{whole} mutually-recursive class definition, which invalidates the
very idea of extensibility. 

Our solution for the problem, again, utilises the idea of open recursion. In short, we parameterise the concrete transformation classes with the same transformations
for \emph{all} components of mutually recursive definition. Since this parameterisation violates the conventions on class interfaces we first generate auxiliary classes.
For our example this auxiliary classes look as follows:

\begin{lstlisting}
   class $\inbr{show\_stub_{expr}}$ $f_{expr}$ $f_{def}$ =
   object inherit [unit, _, string] $\inbr{expr}$ $f_{expr}$
     $\dots$
     method $\inbr{LocalDef}$ $\iota$ x d e = $\dots$ ($f_{def}$ $\dots$) $\dots$
   end
     
   class $\inbr{show\_stub_{def}}$ $f_{expr}$ $f_{def}$ =
   object inherit [unit, _, string] $\inbr{def}$ $f_{def}$
     method $\inbr{Def}$ $\iota$ x s e = $\dots$ ($f_{expr}$ $\dots$) $\dots$
   end
\end{lstlisting}

Note the absence of mutually recursive class definitions. Then, we generate a fix point operator for each mutually recursive definition:

\begin{lstlisting}
   let $\inbr{fix_{expr, def}}$ ($c_{expr}$, $c_{def}$) =
     let rec $t_{expr}$ $\iota$ x = $\inbr{gcata_{expr}}$ ($c_{expr}$ $t_{expr}$ $t_{def}$) $\iota$ x
     and $t_{def}$ $\iota$ x = $\inbr{gcata_{def}}$ ($c_{def}$ $t_{expr}$ $t_{def}$) $\iota$ x in
     ($t_{expr}$, $t_{def}$)
\end{lstlisting}

Here $c_{expr}$ and $c_{def}$ are object generators which take the transformation functions for all components of mutually recursive definition
as parameters. Note, the same fix point generator can be used to construct any concrete transformation for given mutually recursive definition.

With auxiliary classes and the fix point operator we can construct the default implementations for any concrete transformation:

\begin{lstlisting}
   let $\inbr{show_{expr}}$, $\inbr{show_{def}}$ =
     $\inbr{fix_{expr,def}}$ (new $\inbr{show\_stub_{expr}}$, new $\inbr{show\_stub_{def}}$) 
\end{lstlisting}

These default implementations, first, are distributed among the typeinfo structures for relevant types and, second, are used to define conventional
transformation classes:

\begin{lstlisting}
   class $\inbr{show_{expr}}$ fself =
   object inherit $\inbr{show\_stub_{expr}}$ fself $\inbr{show_{def}}$ end

   class $\inbr{show_{def}}$ fself =
   object inherit $\inbr{show\_stub_{def}}$ $\inbr{show_{expr}}$ fself end
\end{lstlisting}

Thus, we again made mutually recursive types indistinguishable from the simple ones (in terms of class interfaces), making it possible to
uniformly generate all transformations with separate compilation support.

On the other hand, in order to extend an existing transformation one needs to inherit from \emph{auxiliary} classes and use the custom fix point operator.
For our previously unsuccessful case the implementation is almost as simple as for the single type definition:

\begin{lstlisting}
   let custom_show, _ =
      $\inbr{fix_{expr,def}}$ ((fun $f_{expr}$ $f_{def}$ ->
                     object inherit $\inbr{show\_stub_{expr}}$ $f_{expr}$ $f_{def}$
                       method $\inbr{Const}$ $\iota$ x n = "a constant"
                     end),
                    new $\inbr{show\_stub_{def}}$) 
\end{lstlisting}

In the actual implementation we generate a memoizing fix point combinator, which follows the same pattern we've described in Section~\ref{memofix}. In addition,
we put the fix point combinator into the typeinfo structure, so, for a type ``\lstinline{t}'' the fix point combinator can be addressed as
``\lstinline{fix(t)}''. End users, however, still need to know the structure of mutually-recursive type definitions in order to use the fix point
combinator properly.

There is one subtlety with our support for mutual recursion: we rely on the property, that adding one function per type is enough to implement open recursion.
However, generally speaking, this is not true: take, for example, the following definition:

\begin{lstlisting}
   type ($\alpha$, $\beta$) a = A of $\alpha$ b * $\beta$ b
   and  $\alpha$ b = X of ($\alpha$, $\alpha$) a
\end{lstlisting}

In the parameters of constructor ``\lstinline{A}'' we have here \emph{different} parameterisations of type ``\lstinline{b}'' and, thus, we would need
\emph{two} functions~--- for ``\lstinline{$\alpha$ b}'' and for ``\lstinline{$\beta$ b}''. However, the type ``\lstinline{a}'' is not regular~--- starting with 
the parameterisation ``\lstinline{($\alpha$, $\beta$) a}'' we can end up with ``\lstinline{($\alpha$, $\alpha$) a}'' and ``\lstinline{($\beta$, $\beta$) a}''.
Thus, we have already ruled such definitions out. In this reasoning we assume that mutually recursive definitions are \emph{essential} in the sense that they
can not be split into separate type declarations (i.e. that every pair of types are mutually ``reachable''). If we replace the second definition in the
example above with, say,

\begin{lstlisting}
   ...
   and $\alpha$ b = int
\end{lstlisting}

then we would end up with a case which is not supported by our framework. However, as types ``\lstinline{a}'' and ``\lstinline{b}'' are actually \emph{not}
mutually recursive, the whole definition can be rewritten, which restores the support.

\subsection{Polymorphic Variants}
\label{pv}

We consider the support for polymorphic variants~\cite{PolyVar,PolyVarReuse} as an important feature of our framework since it complements the ability of defining
composable data structures with the ability of creating composable transformations. The main difference between polymorphic variants and usual algebraic
data types is that it is possible to \emph{extend} previously declared polymorphic variants by adding more constructors or to combine a few types into the one. 

Our goal is to provide a \emph{seamless} integration of generic features: when a few types are being combined we would want to acquire all generic
features for the result type by inheriting the same features from the constituent types.

As we said previously, an extra type parameter ``$\epsilon$'' is inferred to an open version of the polymorphic variant type. Thus, the same generic transformation
function can be used to transform a value using a transformation object for a \emph{wider} type\footnote{We refrain from calling this type a ``subtype'' since there is
no subtyping in \textsc{OCaml}.}. This is achieved by a specific form of generic transformation function, which performs an ``opening'':

\begin{lstlisting}
   let $\inbr{gcata_t}$ $\omega$ $\iota$ subj =
     match subj with
     $\dots$
     | C $\dots$ -> $\omega$#$\inbr{C}$ $\iota$ (match subj with #t as subj -> subj) $\dots$
     $\dots$
\end{lstlisting}

This results in applying the methods of transformation object to an opened version of the type, while the transformation function itself still operates only
of the closed version.

When a few polymorphic variant types are combined, the transformation function simply matches a value against type patterns and dispatches the
transformation to the transformation functions of a corresponding constituent type.

%% file: examples.tex
\section{Examples}

In this section we present some examples, written with the aid of our library. In this examples we will use \cd{camlp5} syntax extension,
although \cd{ppxlib} plugin can be used equally. As we said, the library is a direct inheritor of our prior work~\cite{TransformationObjects}, and
all examples from that paper can be implemented using the new version. Here we show some more.

\subsection{Typed Logic Values}

The first example arose in the context of our work on strongly typed logical DSL for \textsc{OCaml}~\cite{OCanren}. One of the
most important construct there was a unification of terms with free logical variables, and dealing with such data structures
involves a lot of tedious and error-prone work. The typical scenario of interaction between a logical and non-logical worlds
is constructing a \emph{goal} containing a data structure with free logical variables and solving it. The solution
provides bindings for these variables, which, in optimistic scenario, do not contain free variables anymore. To construct
a goal one would need a systematic way to introduce logic variables in some typed data structure, and to recover answers~---
a systematic way to return to a plan, non-logical representation.

The (simplified) type for logic values can be defined as follows:

\begin{lstlisting}
   @type 'a logic =
   | V     of int
   | Value of 'a
   with show, gmap
\end{lstlisting}

A logic value can either be a free logic variable (``\lstinline{V}'') or a some other value (``\lstinline{Value}'') which is not
a free variable (but which can possibly contain free variables inside). To convert to- and from- the logic domain we can use the following
functions:

\begin{lstlisting}
   let lift x = Value x
  
   let reify  = function
   | V     _ -> invalid_arg "Free variable"
   | Value x -> x
\end{lstlisting}

The function ``\lstinline{reify}'' raises and exception on a free variable; indeed, if an occurrence of a free variable
is encountered the logic value can no longer be considered as a regular (non-logical) data structure and has to be interpreted
in some other way.

When we dealing with logic data structures we need to have an opportunity to put a free variable in an arbitrary
position. This means that we have to switch to another type, ``lifted'' into the logic domain. For example,
for arithmetic expressions, which we use as an example through the paper, we would need to construct a value like

\begin{lstlisting}
   Value (
     Binop (
       V 1, 
       Value (Const (V 2)),
       V 3
    )
   )
\end{lstlisting}

which has a type ``\lstinline{lexpr}'', defined as

\begin{lstlisting}
   type expr' = Var of string logic | Const of int logic | Binop of lexpr * lexpr
   and  lexpr = expr' logic
\end{lstlisting}

We also need to implement two conversion functions. All these definitions present a typical example of boilerplate code.

With our framework the solution is almost purely declarative\footnote{But we need to switch the compiler into \cd{-rectypes} mode}.
First, we abstract the type of interest, replacing all positions, in which we may desire to place a type variable, with
fresh type parameters:

\begin{lstlisting}
   @type ('string, 'int, 'expr) a_expr =
   | Var   of 'string
   | Const of 'int
   | Binop of 'string * 'expr * 'expr with show, gmap
\end{lstlisting}

Here we abstract the type of everything, but we could equally abstract it only of itself. Note, we make use of two
generic features~--- ``\lstinline{show}'' and ``\lstinline{gmap}''. The first one is needed for debugging purposes, while
the second is essential for our solution.

Now we can define the logical and non-logical counterparts as customised versions of the abstracted type:

\begin{lstlisting}
   @type expr  = (string, int, expr) a_expr with show, gmap
   @type lexpr = (string logic, int logic, lexpr) a_expr logic with show, gmap
\end{lstlisting}

Note, the ``new'' type ``\lstinline{expr}'' is equivalent to the ``old'' one, thus, this transformation makes no
harm to the existing code.

Finally, the definitions of conversion functions make use of the generic ``\lstinline{gmap}'' feature the
framework provides:

\begin{lstlisting}
   let rec to_logic   expr = gmap(a_expr) lift  lift  to_logic  expr
   let rec from_logic expr = gmap(a_expr) reify reify from_logic @@ reify expr
\end{lstlisting}

As we can see, the support for type constructor application is vital for the success of this scenario. In our prior
implementation~\cite{TransformationObjects} type constructor application was not supported and could not be easily added.

\subsection{Conversion to a Nameless Representation}

Polymorphic variant types make it possible to define composable statically typed and separately compiled data structures~\cite{PolyVarReuse}.
Dealing with them to implement composable statically typed and separately compiled transformations looks like a natural idea. The problem of
constructing transformations from separately compiled, strongly typed components is known as ``The Expression Problem''~\cite{ExpressionProblem}, which
is often used as a ``litmus test'' for generic programming frameworks~\cite{ObjectAlgebras,ALaCarte}. In this section we show the solution for
the expression problem with the aid of our framework. For a concrete problem we take the transformation from named to a nameless representations
for lambda terms.

First, we define the non-binding part of the terms:

\begin{lstlisting}
   @type ('name, 'lam) lam = [
   | `App of 'lam * 'lam
   | `Var of 'name
   ] with show
\end{lstlisting}

Separating this type looks a natural idea since potentially there can be many binding constructs (lambdas, lets, etc.) and by combining them
with the non-binding part (and with themselves) one can acquire a variety of languages with a coherent behaviour.

The type ``\lstinline{lam}'' is polymorphic: the first parameter is used to represent names or de Bruijn indices, the second one is needed
for open recursion (we here follow the known technique for describing extensible data structures with polymorphic variants~\cite{PolyVarReuse}).

What would the transformation to the nameless representation look like for this type? In our terms, what the transformation class is? It is shown
below:

\begin{lstlisting}
   class ['lam, 'nameless] lam_to_nameless
     (flam : string list -> 'lam -> 'nameless) =
   object
     inherit [string list, string, int,
              string list, 'lam, 'nameless,
              string list, 'lam, 'nameless] $\inbr{lam}$
     method $\inbr{App}$ env _ l r = `App (flam env l, flam env r)
     method $\inbr{Var}$ env _ x   = `Var (index env x)
   end
\end{lstlisting}

First, we use a list of strings as an environment, and we pass it as an inherited attribute. Then, we use a function ``\lstinline{index}'' to find a
position of a string in the environment (thus, it translates names to the de Bruijn indices). The interesting part is the typing of the common ancestor
class ``$\inbr{lam}$''. The first triple of its parameters describes the transformation for the first type parameter of the type. As we can see, we
transform strings into integers, using an environment. Next, the type variable ``\lstinline{'lam}'', as we know, will be set to the open version of the ``\lstinline{lam}''.
Finally, the result of the transformation is typed as ``\lstinline{'nameless}''. This is because the result will be, indeed, a different type, as we
will see soon. As the type parameter ``\lstinline{'lam}'' designates the type itself, the last three parameters repeat the next to last three.

Now we define a binding construct~--- abstraction:

\begin{lstlisting}
   @type ('name, 'lam) abs = [ `Abs of 'name * 'lam ] with show
\end{lstlisting}

The same reasoning applies here: we use an open recursion and a parameterization over name representation. The transformation class can be
implemented in a similar manner:

\begin{lstlisting}
  class ['lam, 'nameless] abs_to_nameless
    (flam : string list -> 'lam -> 'nameless) =
  object
    inherit [string list, string, int,
             string list, 'lam, 'nameless,
             string list, 'lam, 'nameless] $\inbr{abs}$
    method $\inbr{Abs}$ env name term = `Abs (flam (name :: env) term)
  end
\end{lstlisting}

Note, the method ``$\inbr{Abs}$'' constructs a value which has a \emph{different} type, than any parameterization of ``\lstinline{abs}''. Indeed, in a
nameless representation abstraction does not keep any name.

We can now combine two type definitions to build a type for terms with binders:

\begin{lstlisting}
   @type ('name, 'lam) term = [ ('name, 'lam) lam | ('name, 'lam) abs) ] with show
\end{lstlisting}

We can also provide two new types for named and nameless representation\footnote{We need to enable \cd{-rectypes} mode for these definitions to compile.}:

\begin{lstlisting}
   @type named    = (string, named) term with show
   @type nameless = [ (int, nameless) lam | `Abs of nameless] with show
\end{lstlisting}

Finally, we build a transformation for converting a named to a nameless representation:

\begin{lstlisting}
   class to_nameless
     (fself : string list -> named -> nameless) =
   object
     inherit [string list, named, nameless] $\inbr{named}$
     inherit [named, nameless] lam_to_nameless fself
     inherit [named, nameless] abs_to_nameless fself
   end
\end{lstlisting}

This transformation is constructed by inheriting all relevant counterparts: a common ancestor class for all transformations for the type ``\lstinline{named}'' and
two concrete transformations for its counterparts. The transformation function can be build in a standard way:

\begin{lstlisting}
   let to_nameless term =
     transform(named) (fun fself -> new to_nameless fself) [] term
\end{lstlisting}

Thus, we constructed a solution for a type from the solutions for its counterparts. This partial solutions can be separately compiled, and the whole
system remains strongly statically typed.

\subsection{A Custom Plugin}
\label{pluginExample}

Finally we demonstrate the utilisation of the plugin system using the example of a fresh custom plugin implementation. For this purpose we
take a well-known \emph{hash-consing} transformation~\cite{HC}. This transformation converts a data structure to its maximally shared
representation, when structurally equal substructures are represented by the same physical object. For example, an expression tree

\begin{lstlisting}
   let t =
     Binop ("+",
       Binop ("-",
         Var "b",
         Binop ("*", Var "b", Var "a")),
       Binop ("*", Var "b", Var "a"))
\end{lstlisting}

can be rewritten into

\begin{lstlisting}
   let t =
     let b  = Var "b" in
     let ba = Binop ("*", b, Var "a") in
     Binop ("+", Binop ("-", b, ba), ba)  
\end{lstlisting}

where equal sub expressions are represented by shared sub trees.

Our plugin for a type ``\lstinline|$\left\{\alpha_i\right\}$ t|'' will provide a hash-consing function ``\lstinline{hc(t)}'' of the type

\begin{lstlisting}
    $\{$ H.t -> $\alpha_i$ -> H.t * $\alpha_i$ $\}$ -> H.t -> $\left\{\alpha_i\right\}$ t -> H.t * $\left\{\alpha_i\right\}$ t
\end{lstlisting}

where ``\lstinline{H.t}''~--- a heterogeneous hash table for values of arbitrary types. The interface for the hash table is
as follows:

\begin{lstlisting}
   module H :
   sig
     type t
     val hc : t -> 'a -> t * 'a
   end
\end{lstlisting}

The function ``\lstinline{H.hc}'' takes a hash table and some value and returns a possibly updated table and a structurally equivalent value
of the same type. For now we postpone the description of this module implementation and consider an example of constructor transformation
method:

\begin{lstlisting}   
   method $\inbr{Binop}$ h _ op l r =
     let h, op = hc(string) h op in
     let h, l  = fself h l in
     let h, r  = fself h r in
     H.hc h (Binop (op, l, r))
\end{lstlisting}

The method takes an inherited attribute~---this time a hash table ``\lstinline{h}'',~--- the whole expressions node (which we do not
need in this case, hence underscore), and three arguments of the constructor: ``\lstinline{op}'' of type \lstinline{string}, and
``\lstinline{l}'' and ``\lstinline{r}'' of type \lstinline{expr}. We first hash-cons all three arguments (which gives us a possibly updated
hash table and three hash-consed values of the same types), then we apply the constructor and hash-cons the value again. To hash-cons
the arguments of the constructor we can use the functions provided by the framework~--- for the type \lstinline{string} it is
``\lstinline{hc(string)}''\footnote{Generally speaking, we would need to implement a hash-consing function for each primitive type; in
  our case, however, we could equally use ``\lstinline{H.hc}''.}, and for both sub expressions it is ``\lstinline{fself}''.

As a final component we need to decide on the type parameters for a plugin class for a type ``\lstinline|$\{\alpha_i\}$ t|''. Clearly,
all inherited attribute types has to be ``\lstinline{H.t}'', and synthesised attribute types has to be ``\lstinline{H.t * $a$}'' for the
type of interest ``$a$''. This gives us the following plugin class definition:

\begin{lstlisting}
   class [$\{\alpha_i\}$, $\epsilon$] $\inbr{hc_t}$ $\dots$ =
   object
     inherit [$\{$ H.t, $\alpha_i$, H.t * $\alpha_i$ $\}$, H.t, $\epsilon$, H.t * $\epsilon$] $\inbr{t}$
     $\dots$
   end
\end{lstlisting}

For simplicity we omitted the specification of functional parameters for the class since their types can be trivially
recovered.

Now we need to generate this logic using a plugin.

The infrastructure code for the plugin implementation is shown below:

\begin{lstlisting}
   let trait_name = "hc"
  
   module Make (AstHelpers : GTHELPERS_sig.S) =
     struct
     
       open AstHelpers

       module P = Plugin.Make (AstHelpers)

       class g tdecls =
       object (self : 'self)
         inherit P.with_inherited_attr tdecls as super
         $\ldots$
       end

     end

   let _ =
     Expander.register_plugin trait_name (module Make : Plugin_intf.Plugin)
\end{lstlisting}

To implement a plugin, one needs to implement a functor parameterised by a helper module, which resembles ``\cd{Ast_builder}'' from
\cd{ppxlib} to create \textsc{OCaml} syntax trees. We need to use a functor since we have to provide two implementations for
a plugin~--- for \cd{camlp5} syntax extension as well as for \cd{ppxlib} itself. The main entity in the body of the functor is
a class ``\lstinline{g}'' declaration (``generator''), which for simplicity can be inherited from one of generic classes 
from the framework. In this case we, first, instantiate the generic plugin ``\lstinline{P}'' for ``\lstinline{AstHelpers}'' and
then inherit from the class ``\lstinline{P.with_inherited_attr}'', which means that we are going to implement a plugin
making use of inherited attribute. The class takes a type declaration as a parameter. Finally, we register the functor as a
first-class module in the framework to make it accessible.

Now we show what the methods of the generator class look like. First, we need to specify what are the types of inherited and
synthesised attributes for the plugin:

\begin{lstlisting}
   method main_inh ~loc _tdecl = ht_typ ~loc

   method main_syn ~loc ?in_class tdecl =
     Typ.tuple ~loc
       [ ht_typ ~loc
       ; Typ.use_tdecl tdecl
       ]

   method inh_of_param tdecl _name =
       ht_typ ~loc:(loc_from_caml tdecl.ptype_loc)

   method syn_of_param ~loc s =
     Typ.tuple ~loc
       [ ht_typ ~loc
       ; Typ.var ~loc s
       ]
\end{lstlisting}

where we assume ``\lstinline{ht_typ}'' is defined as

\begin{lstlisting}
   let ht_typ ~loc =
     Typ.of_longident ~loc (Ldot (Lident "H", "t"))
\end{lstlisting}

In other words, we say here that the type of inherited attribute is always ``\lstinline{H.t}'' and the type of a synthesised attribute for
a type of interest ``\lstinline{t}'' is ``\lstinline{H.t * t}''.

The next group of methods specifies the behaviour of plugin class type parameters:

\begin{lstlisting}
   method plugin_class_params tdecl =
     let ps =
       List.map tdecl.ptype_params ~f:(fun (t, _) -> typ_arg_of_core_type t)
     in
     ps @
     [ named_type_arg ~loc:(loc_from_caml tdecl.ptype_loc) @@
       Naming.make_extra_param tdecl.ptype_name.txt
     ]

   method prepare_inherit_typ_params_for_alias ~loc tdecl rhs_args =
     List.map rhs_args ~f:Typ.from_caml
\end{lstlisting}

The first method specifies the type parameters for the plugin class itself: this time they are exactly the type parameters of the type declaration plus
the extra parameter ``$\epsilon$''. The second one describes the method of recalculation of type parameters for application of type constructor: when
the type declaration looks like

\begin{lstlisting}
   type $\{\alpha_i\}$ t = $\{a_i\}$ tc
\end{lstlisting}

we need to acquire the implementation of the plugin for ``\lstinline{t}'' from the implementation of the same plugin for ``\lstinline{tc}'', inheriting
from properly instantiated corresponding class. As for our plugin the class is parameterised by the same types as the type, we just keep the parameters.

The last group of methods generate the bodies of constructor transformation. As we support regular constructors with both tuple and record
argument specifications as well as top-level tuples and records, there are four methods, which as a rule share many details of implementation. We show the
skeleton for one of them:

\begin{lstlisting}
method on_tuple_constr ~loc ~is_self_rec ~mutual_decls ~inhe tdecl constr_info ts =
  $\dots$ 
  match ts with
  | [] -> Exp.tuple ~loc [ inhe; c [] ]
  | ts ->
     let res_var_name = sprintf "%s_rez" in
     let argcount = List.length ts in
     let hfhc =
       Exp.of_longident ~loc (Ldot (Lident "H", "hc"))
     in
     List.fold_right
       (List.mapi ~f:(fun n x -> (n, x)) ts)
       ~init:$\dots$
       ~f:(fun (i, (name, typ)) acc ->
            Exp.let_one ~loc
              (Pat.tuple ~loc [ Pat.sprintf ~loc "ht%d" (i+1)
                              ; Pat.sprintf ~loc "%s" @@ res_var_name name])
              (self#app_transformation_expr ~loc
                 (self#do_typ_gen ~loc ~is_self_rec ~mutual_decls tdecl typ)
                 (if i = 0 then inhe else Exp.sprintf ~loc "ht%d" i)
                 (Exp.ident ~loc name)
              )
              acc
          )
  $\dots$
\end{lstlisting}

This implementation makes use of the generic method ``\lstinline{self#app_transformation_expr}'' from the framework, which generates an application of
the transformation in question for a given type.

The final component for the implementation is module ``\lstinline{H}'' itself. The standard functor ``\lstinline{Hashtbl.Make}'' instantiates a
hash table making use of some hash function and equality predicate, supplied by an end user. In a whole, we follow a conventional pattern:
for the hash function we use polymorphic ``\lstinline{Hashtbl.hash}'' and for the equality we use physical equality ``\lstinline{==}''. There are, however, two
subtleties:

\begin{itemize}
\item Since our hash table is heterogeneous, we have to utilise unsafe coercion ``\lstinline{Obj.magic}''.
\item Our implementation for equality has to be a little more complex than simple ``\lstinline{==}'': we need to compare the top-level constructors and
  the number of their arguments \emph{structurally}, and only then compare the corresponding arguments by physical equality. Technically this
  may result in hash-consing structurally equal values of \emph{different} types.
\end{itemize}

We rely here on the follow observation: as hash-consing is only consistent with referentially-transparent data structures, we can assume
that structurally equal data structures can be interchangeable regardless their types. The complete implementation for this plugin can be seen in the main project
repository; it occupies 164 LOC, including comments and blank lines.

%% file: related.tex
\section{Related Works}

As our work makes use of both functional (combinators) and object-oriented (classes and objects) features of \textsc{OCaml} there are some relevant works
in both domains of typeful functional and object-oriented programming. The most relevant framework, developed for \textsc{OCaml}, which utilises the same
ideas, but makes essentially different design decisions, is \textsc{Visitors}~\cite{Visitors}; we postpone the in-depth comparison of our framework with
\textsc{Visitors} until the end of this section.

First, there is a number of frameworks for generic programming in \textsc{OCaml}, which utilise a completely generative approach~\cite{Yallop,PPXLib}~---
all requested generic functions for all types are generated by the framework separately. This approach is very practical as long as the assortment
of shipped functions is rich enough and sufficient for a given use case. However, if not, someone has to extend the framework, implementing
all missing functions anew (and, potentially, with a very little code reuse). In addition, the functions themselves are hard coded and
lack extensibility. With our framework, first, many end-user generic functions can be easily derived from the generated ones, and second, in order to
implement a completely fresh plugin it is sufficient to hard code only ``the interesting'' part, as the generation of the single traversal
function and transformation class are already provided by the framework itself.

A number of approaches to functional generic programming utilises the idea of type \emph{representation}~\cite{Hinze}.
The idea is to develop a uniform representation for any type under transformation and provide two conversion functions from- and to this representation
(ideally, comprising an isomorphism). A generic function performs transformation on a representation of actual data structure, which makes it possible to
implement every such function only once. The conversion functions themselves can in turn be constructed (semi) automatically using such features of
the language type system as type classes~\cite{Hinze,ALaCarte} or type families~\cite{InstantGenerics} (in \textsc{Haskell}) or generated using syntax extension
mechanism~\cite{GenericOCaml} (in \textsc{OCaml}). While some of these approaches allow extension and modification of generic functions by, for example, specifying a
specific treatment for some types or supporting extensible types, our solution is still more flexible as it allows modification with granularity of individual
constructors. In addition, with our framework it is possible for multiple versions of the same generic function for the same type to coexist.

A different approach is taken in ``Scrap Your Boilerplate'', or SYB~\cite{SYB}, initially developed for \textsc{Haskell}. This approach makes it
possible to implement transformations which identify the occurrences of instances of a certain datatype inside arbitrary data structure. Two main
kinds of transformations are supported: \emph{queries}, which collect and return the instances of the designated datatype based on some user-defined
criterion, and \emph{transformations}, which uniformly propagate some type-preserving transformation for a datatype of interest. In the follow-up papers
the approach was extended to deal with transformations which traverse pairs of data structures~\cite{SYB1} and to support the extension of already implemented
transformations with new type cases~\cite{SYB2}. Later the approach was implemented  for other languages, including \textsc{OCaml}~\cite{SYBOCaml,Staged}.
Unlike our case, SYB takes the route of discriminating on a whole type, not individual constructors. In addition the shape of available transformations look rather
restrictive, and, once implemented, transformations for a given type can not be modified. It is interesting, that, potentially, SYB-style generic functions
can ``break through the encapsulation barrier''~--- indeed, they can identify the occurrences of values of type of interest inside \emph{arbitrarily typed}
data structures. Thus, their behaviour depend on the actual details of data structure organisation, including those which were intentionally hidden by encapsulation.
This may result in, first, the possibility for undesirable reverse-engineering (by applying various type-sensitive transformations and analysing the results) and,
second, in fragility of interfaces~--- after a modification of data structure implementation generic functions for \emph{old} version can still be applied with
neither static nor dynamic error, but with wrong (or undesirable) results. 

There is a certain similarity between our approach and \emph{object algebras}~\cite{ObjectAlgebras}. Object algebras were proposed as a solution
for expression problem in mainstream object-oriented languages (\textsc{Java}, \textsc{C++}, \textsc{C\#}), which would not require advanced type system features besides
regular inheritance and generics. In the original exposition object algebras were presented as a design and implementation pattern; the follow-up
works have improved the initial proposal in various directions~\cite{ObjectAlgebrasAttribute,ObjectAlgebrasSYB}.
With object algebras a data structure under transformation is also encoded using the method-per-variant (constructor) idea, which makes it possible to
provide the extensibility in both dimensions and retroactive implementation. However, being developed for essentially different language environment,
the solution using object algebras would differ from ours in many concrete aspects. First, with object algebras the ``shape'' of a data structure has to
be represented by a generic function, which takes a concrete object algebra instance as a parameter (``Church encoding'' for types~\cite{Hinze}). Applying
this function to various implementations of object algebra one can acquire various transformations (for example, printing). To instantiate the data
structure itself one needs to provide a specific object algebra instance~--- \emph{factory}. However, after the instantiation the data structure itself
can not be generically transformed anymore. Thus, object algebras force end users to switch to data-as-function representation, which may or may not be
beneficial in different concrete cases. In contrast our approach non-destructively adds new functionality to the familiar world of algebraic data types,
pattern matching and recursive functions. Generic transformation implementations are completely separated from data representation, and end users may
freely transform their data structures in a familiar way without losing the ability to apply (or extend) generic functions. Another difference stems
from the fact that in our case, unlike mainstream object-oriented languages, polymorphic variants are used as a main tool for datatype extension.
Supporting polymorphic variants as a mean for datatype extensibility requires a fresh solution.

Finally, among existing generic programming frameworks for \textsc{OCaml} we can name two, which resemble ours: \cd{ppx\_deriving}/\cd{ppx\_traverse}
(a part of \cd{ppxlib}~\cite{PPXLib}) and \textsc{Visitors}~\cite{Visitors}.

\cd{ppx\_deriving} is the simplest approach possible: type declarations are mapped one-to-one to recursive functions representing a specific kind of
transformation. It is the most efficient implementation (functions are called directly, no late biding involved) but it is not extensible. If end users
need to modify slightly the generated function, they should copy and paste generated code to modify it manually. The amount of work to support a new
transformation will drastically increase if type definitions change during the development cycle.

In \cd{ppx\_traverse} extensible transformations are represented as objects; unlike our case, method-per-type approach is used. In addition 
\cd{ppx\_traverse} does not make use of inherited attributes, thus some transformations like equality or comparison are not representable.

\textsc{Visitors}, on the other hand, explores a similar to ours object-oriented approach, in which many decisions, rejected by us, were taken (and vice versa). Here
we summarise the main differences:

\begin{itemize}
   \item \textsc{Visitors} is excessively object-oriented~--- in order to use it one needs to instantiate some object and call proper method. In our case as long as
     only predefined features are required one can use a more native combinatorial interface.
     
   \item \textsc{Visitors} implements a number of useful transformations in an \emph{ad-hoc} manner; in our case all transformations are instances of the
     same generic scheme. It is possible to combine different transformations via inheritance as long as the types of underlying scheme unify. We also argue, that
     in our framework the implementation of user-defined plugins is much easier.
     
   \item Following SYB, \textsc{Visitors} takes a type-discriminating route: for each type of interest (including the built-in ones) there is a dedicated
     transformation method in each object, representing a transformation. While this solution indeed adds some flexibility, we firmly oppose it, since it
     breaks the encapsulation: inspecting the methods of a transformation (which cannot be hidden in a module signature) one can retrieve some
     information about the implementation of encapsulated types. Even worse, the data structures of abstract types can be manipulated in an unprescribed
     manner using the public type-transforming interface.

   \item In our case the type parameters for transformation classes have to be specified by an end user. With \textsc{Visitors} this burden is offloaded to the
     compiler with the aid of some neat trick. However, this trick makes it impossible to use \textsc{Visitors} syntax extension in module signatures. There is no
     such problem in our case~--- our framework can be equally used in both implementation and interface files.

   \item \textsc{Visitors} in its current state\footnote{The latest available version is 20180513} does not support polymorphic variants.
   
   \item \textsc{GT} supports arbitrary type constructor applications but \textsc{Visitors} in its current state doesn't (both in monomorphic and polymorphic mode).
     For instance, the following example doesn't compile:
     
   \begin{lstlisting}
      type ('a,'b) alist = Nil | Cons of 'a * 'b
      [@@deriving visitors { variety = "map"; polymorphic = true }]

      type 'a list = ('a, 'a list) alist
      [@@deriving visitors { variety = "map"; polymorphic = false }]
   \end{lstlisting}
   
   Moreover, adding an extra construct doesn't solve the problem:
   
    \begin{lstlisting}
       type 'a list = L of ('a, 'a list) alist [@@unboxed]
       [@@deriving visitors { variety = "map"; polymorphic = false }]
    \end{lstlisting}
    
    There is also an issue with type aliases in polymorphic mode (monomorphic part of \textsc{Visitors} compiles successfully):
    
    \begin{lstlisting}
       type ('a,'b) t = Foo of 'a * 'b (* OK *)
       [@@deriving visitors { variety = "map"; polymorphic = true }]
       
       type 'a t2 = ('a, int) t
       [@@deriving visitors { variety = "map"; name="yyy"; polymorphic = true }]
    \end{lstlisting}
    
    The generated code can be fixed manually by removing explicit polymorphic type annotations from objects' methods, which leads to the code
    very similar to the one generated by \textsc{GT}. From these we can conclude that \textsc{GT} can be seen an a reimplementation of polymorphic
    mode of \textsc{Visitors} where more type declarations compile successfully.
    
\end{itemize}

%% file: conclusion.tex
\section{Future Work}

There are a few possible directions for future work. First, in this paper we did not address the performance issues. As we represent
the transformations in a very generic form with many levels of indirection, obviously, the transformations, implemented with
our framework, are at disadvantage in comparison with hard coded ones in terms of performance. We assume that the performance of transformations
can be essentially improved by applying some techniques like staging~\cite{Staged} or, perhaps, object-specific optimisations.

Another important direction is supporting more kinds of type declarations, in the first hand, GADTs and non-regular types. Although we have some
implementation ideas for this case, the solution we came up with so far makes the interface of the whole framework too cumbersome to use even for
simple cases.

Finally, the typeinfo structure we generate can be used to mimic the \emph{ad-hoc} polymorphism as it contains the implementation of
type-indexed functions. This, together with some proposed extensions~\cite{ModularImplicits}, can open interesting perspectives.